\documentclass[reprint,aps,pra,superscriptaddress,longbibliography]{revtex4-1}
\usepackage[latin9]{inputenc}
\usepackage{color}
\usepackage{times}
\usepackage{amsmath}
\usepackage{amssymb}
\usepackage{stmaryrd}
\usepackage{graphicx}
\usepackage[unicode=true,
 bookmarks=true,bookmarksnumbered=false,bookmarksopen=false,
 breaklinks=false,pdfborder={0 0 1},backref=false,colorlinks=true]
 {hyperref}
\hypersetup{linkcolor=magenta, urlcolor=blue, citecolor=blue, pdfstartview={FitH}, hyperfootnotes=true, unicode=true}
 
\makeatletter
\@ifundefined{textcolor}{}
{%
 \definecolor{BLACK}{gray}{0}
 \definecolor{WHITE}{gray}{1}
 \definecolor{RED}{rgb}{1,0,0}
 \definecolor{GREEN}{rgb}{0,1,0}
 \definecolor{BLUE}{rgb}{0,0,1}
 \definecolor{CYAN}{cmyk}{1,0,0,0}
 \definecolor{MAGENTA}{cmyk}{0,1,0,0}
 \definecolor{YELLOW}{cmyk}{0,0,1,0}
}

\newcommand{\doublewidetilde}[1]{{%
  \mathpalette\double@widetilde{#1}%
}}
\newcommand{\double@widetilde}[2]{%
  \sbox\z@{$\m@th#1\widetilde{#2}$}%
  \ht\z@=.9\ht\z@
  \widetilde{\box\z@}%
}

\usepackage{amsfonts}\usepackage{tabularx}\usepackage{dcolumn}\usepackage{bm}\usepackage{graphicx}\usepackage{epstopdf}

\hypersetup{urlcolor=blue}

\usepackage{tensor}
\usepackage{braket}

\usepackage[capitalise,compress]{cleveref}
\crefname{section}{Sec.}{Secs.}
\Crefname{section}{Section}{Sections}
\crefrangelabelformat{equation}{\textup{(#3#1#4)}--\textup{(#5#2#6)}}

\makeatother

\begin{document}

\newcommand{\SG}[1]{\textcolor{magenta}{\textbf{#1}}}
\newcommand{\YW}[1]{\textcolor{green}{\textbf{#1}}}
\newcommand{\TP}[1]{\textcolor{blue}{\textbf{#1}}}
\newcommand{\JMT}[1]{\textcolor{red}{\textbf{#1}}}

\title{Beyond Spontaneous Emission: Giant Atom Bounded in Continuum}

\author{Shangjie Guo}
\affiliation{Joint Quantum Institute, University of Maryland, College Park, Maryland 20742, USA}
\affiliation{Joint Center for Quantum Information and Computer Science, NIST/University of Maryland,
College Park, Maryland 20742, USA}

\author{Yidan Wang}
\affiliation{Joint Quantum Institute, University of Maryland, College Park, Maryland 20742, USA}
\author{Thomas Purdy}
\affiliation{National Institute of Standards and Technology, Gaithersburg, MD 20899, USA}
\affiliation{Pittsburgh Quantum Institute, University of Pittsburgh, Pittsburgh, PA 15260, USA}

\author{Jacob Taylor}
\affiliation{Joint Quantum Institute, University of Maryland, College Park, Maryland 20742, USA}
\affiliation{Joint Center for Quantum Information and Computer Science, NIST/University of Maryland,
College Park, Maryland 20742, USA}
\affiliation{National Institute of Standards and Technology, Gaithersburg, MD 20899, USA}%

\date{\today}

\begin{abstract}
The quantum coupling of individual superconducting qubits to microwave photons leads to remarkable experimental opportunities. Here we consider the phononic case where the qubit is coupled to an electromagnetic surface acoustic wave antenna that enables supersonic propagation of the qubit oscillations. This can be considered as a giant atom that is many phonon wavelengths long. We study an exactly solvable toy model that captures these effects, and find that this non-Markovian giant atom has a suppressed relaxation, as well as an effective vacuum coupling between a qubit excitation and a localized wave packet of sound, even in the absence of a cavity for the sound waves. Finally, we consider practical implementations of these ideas in current surface acoustic wave devices.
\end{abstract}
\maketitle

\section{Introduction}

    The coupling of resonant, compact systems to continuous media has a rich history, underlying phenomena ranging from musical instruments to complex machinery to the spontaneous emission of light from an atom \cite{SE1,SE2}. The strong coupling regime of such systems has also led to a plethora of applications in cavity quantum electrodynamics (QED) \cite{cavityQED}, circuit QED \cite{circuitQED,circuitwaveguideQED}, and waveguide QED \cite{circuitwaveguideQED,waveguideQED1,waveguideQED2}, all of which work in the regime where light propagation is fast relative to appropriate coupling time scales such as the coherence time. However, collective effects, such as Dicke superradiance, have shown that pre-existing coherence across multiple wavelengths of the medium excitations can dramatically alter the simple dynamics such open quantum systems \cite{Dicke1,Dicke2}.

    Here we examine an example of such long-range coherence in the form of a superconducting qubit coupled nonlocally coupled to a long, quasi-1D phononic waveguide. This system can be realized in, for example, surface acoustic wave (SAW) devices \cite{SAWdevicebook}. Working in the lumped element limit, the electrical antennae that couple to the mechanical waveguide have practically simultaneous coupling to distant regions of the system, while the motional degrees of freedom are constrained to propagate at the speed of sound. This leads to a variety of supersonic phenomena in the quantum acousto-dynamics (QAD) regime which has been heretofore largely unexplored.

    Pioneering work in this domain have labeled this the ``giant atom'' regime of SAW devices \cite{LZGuo2019Exp,Anton2018,LZGuo2017}. This model breaks locality in the lumped element limit and inevitably becomes non-Markovian, requiring a more detailed theoretical treatment \cite{kanu,yidan,nonM1,nonM2,nonM3,nonM4,nonM-BIC,nonMopenrev}. Furthermore, recent experiments show the robustness of systems that couple mechanical with electromagnetic parts in the quantum regime and open the opportunity to realize giant atoms in experiments \cite{LZGuo2019Exp,sletten2019,sletten2018,Cleland,Noguchi2018,Noguchi2017,Anton2017,Anton2015,Anton2014}.
    
    \begin{figure}[t]
        \includegraphics[width=\linewidth]{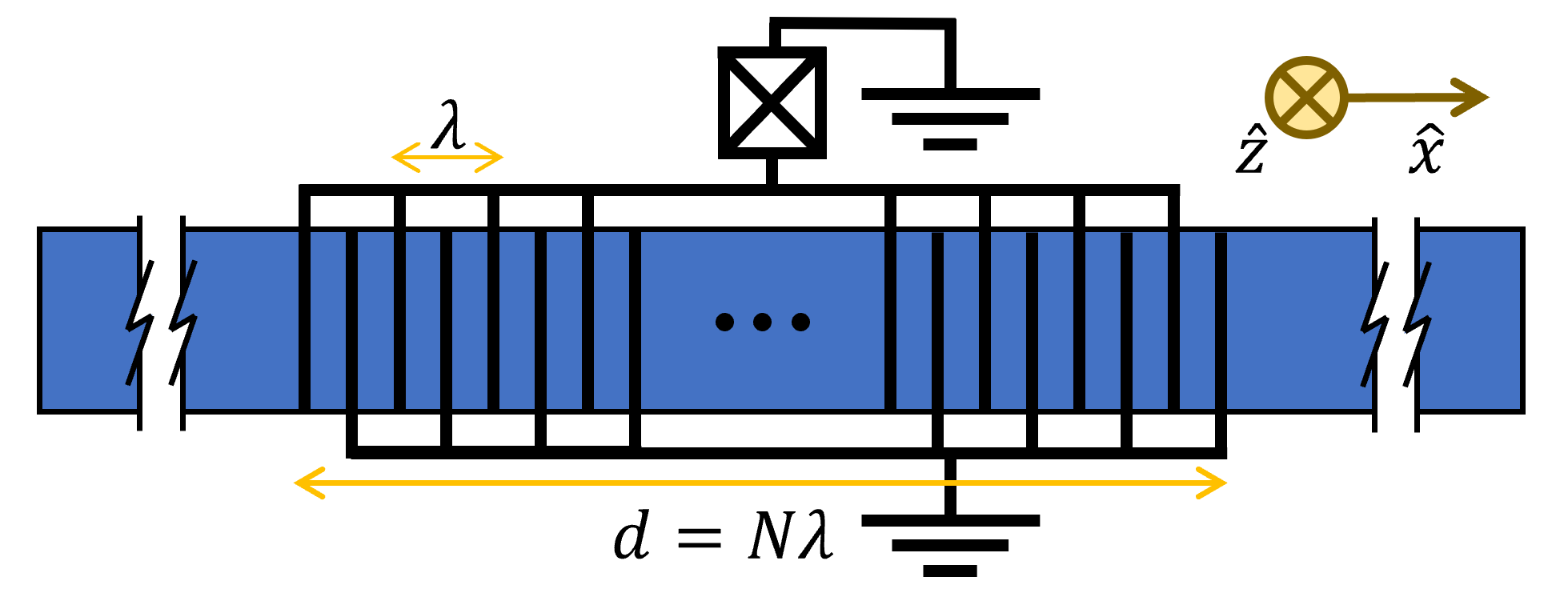}
        \caption{A sketch of a circuit QAD device viewed from the top. Black lines show electrodes and the blue area shows the surface of piezoelectric material substrate. The substrate extends deeply in $+z$ direction.}
        \label{fig:1}
    \end{figure}
    
    We show that these devices have remarkable properties, particularly that of strong coupling without the presence of a cavity, in which a long-lived atomic excitation dynamic emerges due to the coupling to the electrical circuit directly, and the formation of long-lived states of sound in the lossy medium. We describe this as the bounded giant atom phenomenon.
    
    While our simple theoretical model predicts this phenomenon directly, a more complicated numerical approach shows specific additional phase matching conditions that must be satisfied for experimental observation of the strong coupling of this emergent of bounded effect to the quantum bit. Furthermore, in this regime, boundary-based damping of the sound exponentially decreases with the atom size, leading to substantial improvements in coherence times. Our study paves the way for compact, circuit QED qubits with Josephson junctions to as a replacement for current microwave resonator-based approaches for quantum computing.
   
    The rest of this paper is organized as follows: In Sect. II, we review the Weisskopf-Wigner theory for spontaneous emission \cite{SE1}, which provides the structure for our model later; throughout the paper we refer to the superconducting qubit with antennae as a giant atom. We calculate the coupling between the artificial atom and phonons of the circuit QAD device, and we simplify it to a Lorentzian toy model in Sect. III. In Sect. IV, we derive our main results from the toy model and compare our results with the numerical simulation. We conclude in Sect. V and show future applications of the general method presented in this paper. 

\section{Background}

\subsection{The General Theory}

   We consider a two-level giant atom with ground state $\ket{g}$ and excited state $\ket{e}$ with a frequency difference $\nu$ that non-locally couples to an infinitely-long 1-D bosonic field, governed by the following Hamiltonian in the rotating wave approximation: 
    \begin{equation}
    \hat{H} = \frac{\nu}{2} \sigma_z + \int 
    dK  \left[\omega(K) \hat{a}^\dagger_K \hat{a}_K +  g(K; N)\left(\hat{\sigma}_+ \hat{a}_K + \text{h.c.}\right)\right],
    \label{eqn:hamiltonian}
    \end{equation}
    where $\hat{\sigma}_+$($\hat{\sigma}_-$), and $\hat{a}_K^\dagger$($\hat{a}_K$) are creation (annihilation) operators for atomic excitation and field, respectively. They satisfy $(\hat{\sigma}_-)^\dagger = \hat{\sigma}_+ = |e\rangle\langle g|$, $\sigma_z = |e \rangle\langle e| - |g \rangle\langle g|$, and $[\hat{a}_K, \hat{a}_{K'}^\dagger] = \delta(K-K')$. $\nu$ is the atomic transition frequency. We assume that the field has a linear dispersion $\omega(K) = c_s |K|$ with the speed of sound $c_s$, for momentum $K$. We set $\hbar = 1$ for simplicity.

    We consider the coupling $g(K; N)$ to depend on the momentum $K$. As the Fourier transform of the position-dependent coupling, it is also  parameterized by the spatial length of the atom $N$. One can expect that the parameter $N$ will change the atom relaxation dynamics via tuning the shape of $g(K; N)$. We shall discuss two different models for $g(K; N)$ in Sect. III.
    
    We denote the vacuum state by $|g,0\rangle$, and limit our system to a single excitation Hilbert subspace with basis states $|e,0\rangle=\hat{\sigma}_+|g,0\rangle$ and $|g,K\rangle=\hat{a}_K^\dagger|g,0\rangle$, such that any time-dependent state can be described as $|\psi(t)\rangle=\alpha (t)|e,0\rangle+\int_{-\infty}^{+\infty} dK \beta_K (t)|g,K\rangle$, where $\alpha(t)$, and $\beta_K(t)$ are time-dependent amplitudes. In a frame rotating with frequency $\nu$, we derive the equations of motion 
    \begin{align}
        \dot{\alpha}(t) &= -2 i\int_{-\infty}^{+\infty} dk g(k; N) \beta_k (t),
        \label{eqn:a}\\
        \dot{\beta}_k(t) &= -i\delta(k) \beta_k (t) - ig(k; N) \alpha (t). 
        \label{eqn:b}
    \end{align}
    Note that as the coupling is real in position space in our case, such that $g(K;N) = g(-K;N)$, so the two branches for $K>0$ and $K<0$ contribute symmetrically and can be merged in Eq.\ \eqref{eqn:a}. The momentum in the rotating frame is redefined as $k = |K| - \nu/c_s$, such that the field frequency becomes $\delta(k) = \omega(K)-\nu = c_s k$ for the near-resonance regime. Then, by taking the Laplace transform from the time domain into the complex frequency domain by $\tilde{\alpha}(s) = \mathcal{L}[\alpha(t)]$, and $\tilde{\beta}_k(s) = \mathcal{L}[\beta_k(t)]$, we get:
    \begin{align}
    s\tilde{\alpha}(s) -\alpha (0) &= -2i\int_{-\infty}^{+\infty} dk g(k; N) \tilde{\beta}_k(s)\ ,
    \label{eqn:sa}\\
     s\tilde{\beta}_k(s) - \beta_k (0) &= -i\delta(k) \tilde{\beta}_k(s) - ig(k; N) \tilde{\alpha}(s)\ .
     \label{eqn:sb}
     \end{align}

    We set $\alpha (0)=1$ and $\beta_k (0)=0$ to investigate the relaxation of an atomic excitation. Then we have $\tilde{\beta}_k (s) = -i g(k; N) \tilde{\alpha} (s)/(s+i\delta(k))$ and the response function $\chi(s)\equiv \tilde{\alpha}(s)/\alpha (0)$ becomes
    \begin{align}
        \chi(s) = \left(s+2\int_{-\infty}^{+\infty}dk \frac{|g(k; N)|^2}{s+i\delta(k)}\right)^{-1}.
        \label{eqn:response}
    \end{align}
    When $g(k; N)$ is an analytic function, we can derive that $\alpha (t) = \mathcal{L}^{-1}[\chi(s)] \alpha (0) = \sum_{n}\text{Res}[\chi(s),s_n]e^{s_n t}$ from the residue theorem and our initial conditions, where $s_n$ is the $n$th pole of $\chi(s)$ that satisfy the equation: $[\chi(s_n)]^{-1}=0$ for $n\in \{1,2,...,n_{\text{max}}\}$. $n_{\text{max}}$ is the number of the poles of $\chi(s)$. Causality confines $s_n$ to be in the left half complex plane or on the imaginary axis, i.e. $\text{Re}(s_n)\leq 0$ \cite{LRTbook}. Note that the inverse Laplace transform requires that the contour path of integration is in the region of convergence of $\chi(s)$. This can be satisfied by integrating Eq.\ \eqref{eqn:response} with the condition $\text{Re}(s)>0$.
    
    Armed with the solution for the poles $s_n$, we describe the atomic relaxation process as a composition of damped oscillation modes with effective vacuum Rabi oscillation frequencies $\text{Im}(s_n)$ and decay rates $-2\text{Re}(s_n)$. In the long-time limit, only the slowest damped modes can survive, and we thus define the long-time relaxation rate as $\gamma \equiv \text{Min}_n[-2\text{Re}(s_n)]$. 
    
    To understand the giant atom relaxation, we study how the poles of response function $s_n$ change according to the atom size $N$. In the next section, we consider a realistic circuit QAD model and a simpler Lorentzian toy model to characterise $g(k; N)$ with $N$ being a changing parameter, and study the response function $\chi(s)$ and its poles.

\subsection{The Weisskopf-Wigner Limit}

    Before moving into the giant atom case, we first review the Weisskopf and Wigner approach to the point-like atom case\cite{SE1}. A point-like atom  couples to all wavelengths emission equally, i.e. $g(k; N) = g_0$, independent of $k$. In this situation,  one can calculate the real part of the equation $[\chi(s)]^{-1}=0$, which results in
    \begin{equation}
        \gamma_1\equiv -2\text{Re}(s_1) = 4\pi |g_0|^2/c_s.
        \label{eqn:small}
    \end{equation}
    This textbook result shows when a point-like atom couples to an 1-D field, the atom decays with its spontaneous emission rate $\gamma_1$. In the giant atom case, we also define $\gamma_1$ as the weak-coupling relaxation rate for a unit cell (e.g., $N = 1$) for later discussion. Now we can proceed and study $g(k; N)$ for the circuit QAD and the toy models that simplify it.

\section{The Circuit QAD and Toy Models}

\subsection{The Circuit QAD Model}
    
    We examine a simplified 1-D model for the circuit QAD device shown in Fig.\ \ref{fig:1}. A circuit QAD device comprises a superconducting artificial atom (as a Josephson junction) and a surface acoustic wave (SAW) cavity. The qubit couples to the cavity via an inter-digital transducer (IDT), where two interlocking comb-shaped arrays of electrodes are fabricated on the surface of a piezoelectric substrate. Such systems have been used to achieve strong coupling, where the vacuum Rabi coupling exceeds dephasing and damping \cite{sletten2018,Cleland,Noguchi2018,Anton2017}. We can map the spatial atom size to the length of the IDT $d$, and the resonance emission wavelength to the IDT characteristic wavelength $\lambda$ (the finger spacing of the IDT). We use the number of fingers of the IDT $N = d/\lambda$ as the atom size parameter for this circuit QAD model.
    
    Since the electromagnetic wave travels about $10^5$ faster than sound through the IDT, we take the lumped element limit for the circuit, and the electronic subsystem can be regarded as a two-level system that interacts with SAW at different positions simultaneously. Notice that this system inevitably becomes non-Markovian under this assumption, thus necessitating our use of the Laplace transform solutions in what follows, rather than more typical quantum optics approximations. We also assume the mass loading of all electrodes to be zero to remove additional mechanical resonances, and we approximate the stationary electric potential within the IDT region to a triangle wave: $V(x) =  (2 V_0/\pi)\arcsin[\sin(\pi x/2\lambda)]$, where  $V_0$ is the voltage applied on the IDT.
    
\begin{figure}[t]
        \includegraphics[width=\linewidth]{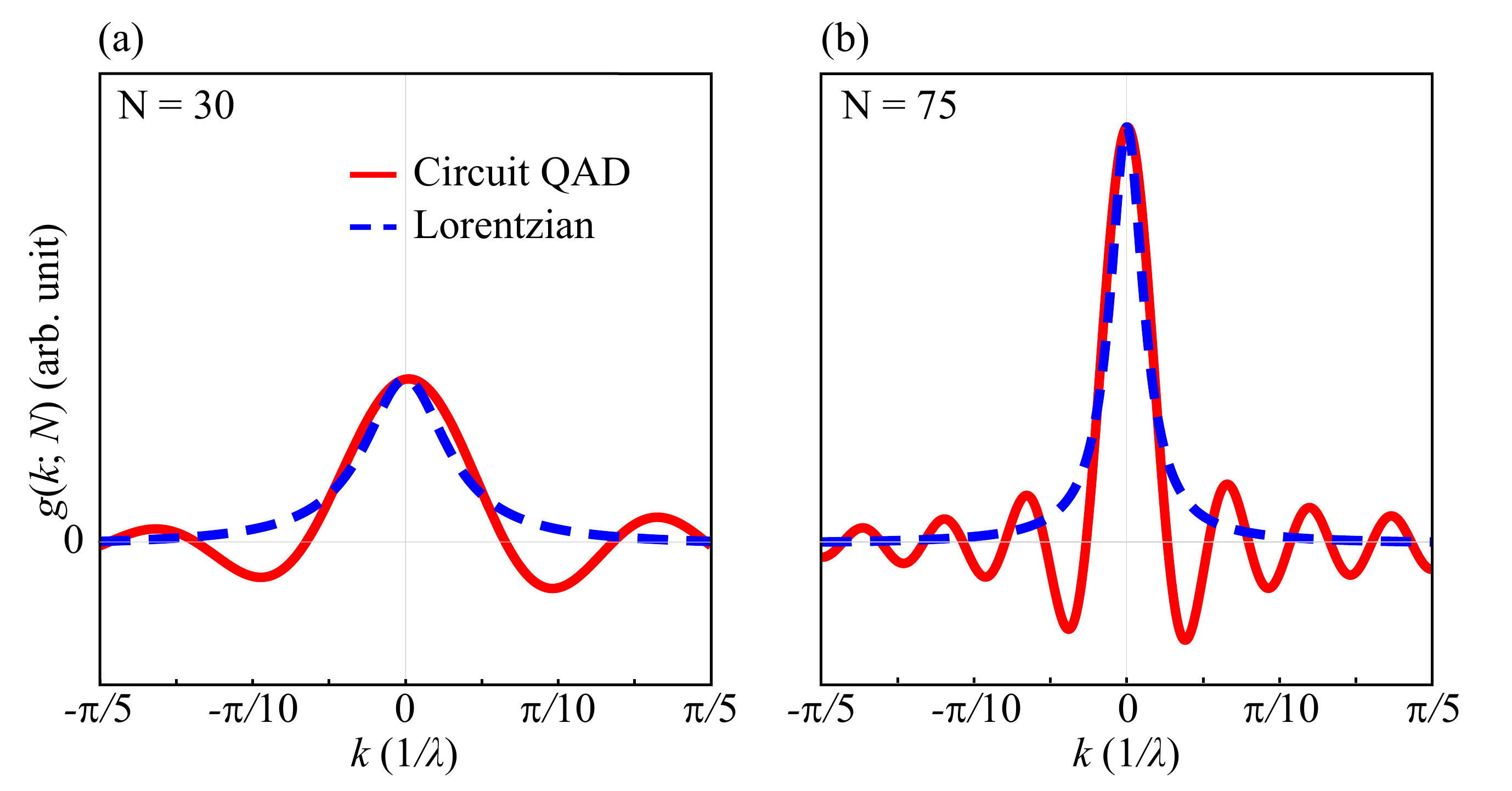}
        \caption{The momentum-dependent coupling $g(k; N)$ for (a) $N = 30$,  (b) $N = 75$. Red solid lines correspond to circuit QAD model \eqref{eqn:cqad}, blue dashed lines to Lorentzian toy model \eqref{eqn:lorentz}. The vertical axes for (a) and (b) share the same scale.}
        \label{fig:2}
    \end{figure}

    We take the atom transition frequency to equal the IDT resonance frequency, i.e. $\nu = 2\pi/T = 2\pi c_s/\lambda$, where $c_s$ is the speed of SAW propagation, and $T$ is the designed fundamental period of the SAW. We calculate the coupling $g(k;N)$ for circuit QAD device as \cite{Schuetz2015}
    \begin{equation}
        g_{\text{cQAD}}(k; N) = \sqrt{\frac{\gamma_1 c_s}{2\pi}}\ \frac{\sin ( N k \lambda /2) \cot(k \lambda /4)}{2+k\lambda/\pi}.
        \label{eqn:cqad}
    \end{equation}
    The derivation of Eq.\ \eqref{eqn:cqad} is given in Appendix A. We illustrate $g_{\text{cQAD}}(k; N)$ in Fig.\ \ref{fig:2} for $N=30$, and $75$. This model has a finite bandwidth about $2\pi/N\lambda$, with the on-resonance coupling proportional to $N$. Note that the poles of the response function \eqref{eqn:response} are hard to find analytically with this model. Therefore, we establish a toy model in the next subsection to capture the long-time dynamics and where we can analytically express its poles. Then, we compare the toy model to numerical results using the circuit QAD model in Sect. IV. B.
    
\subsection{The Lorentzian Toy Model}

    To evaluate the integral in Eq.\ \eqref{eqn:response}, we use a Lorentizian toy model $g_{\text{Lor}}(k; N)$ defined as 
    \begin{equation}
        g_{\text{Lor}}(k; N) \equiv \sqrt{\frac{\gamma_1 c_s}{2\pi}}\ \frac{N}{(N k\lambda/\pi)^2+1},
        \label{eqn:lorentz}
    \end{equation}
    instead of Eq.\ \eqref{eqn:cqad}. Such a model satisfies the following criteria: it has a finite bandwidth about $2\pi/N\lambda$ and an on-resonance coupling proportional to $N$, it is non-local in position with the scale of $N\lambda$, and it decays exponentially in position and quadratically in momentum. In Fig.\ \ref{fig:2}, we illustrate that the shape of the Lorentzian toy model matches the central peak of the circuit QAD model, while it does not capture the oscillation behavior at large $|k|$. This toy model greatly simplifies the calculations and allows us to analytically describe the poles of the response function $\chi(s)$, leading to our main results in Sect. IV. A. We can then analyze corrections to this model from the QAD picture.

\section{Results}

\subsection{Analytic Solutions from the Lorentzian Model}
    
    First, we substitute Eq.\ \eqref{eqn:lorentz} into the equation defining the poles of the response function, $[\chi(s_n)]^{-1}=0$, which yields
    \begin{equation}
        s_n + \frac{N^2 \gamma_1 \nu(N s_n  + \nu )}{(\nu +2N s_n )^2} = 0.
        \label{eqn:main}
    \end{equation}
    This equation can be reduced to a cubic polynomial of $s_n$, and we give the explicit form of its solutions in Appendix B. In Fig.\ \ref{fig:3}(a-b), we set $\gamma_1= \pi\times 10^{-5}\nu$ and plot the $-2\text{Re}(s_n)$ and $\text{Im}(s_n)$, which indicate the damping rates and the effective Rabi frequencies. We mark the solutions associated with the slowest damped modes with solid lines.
    
    \begin{figure}[t]
        \includegraphics[width=\linewidth]{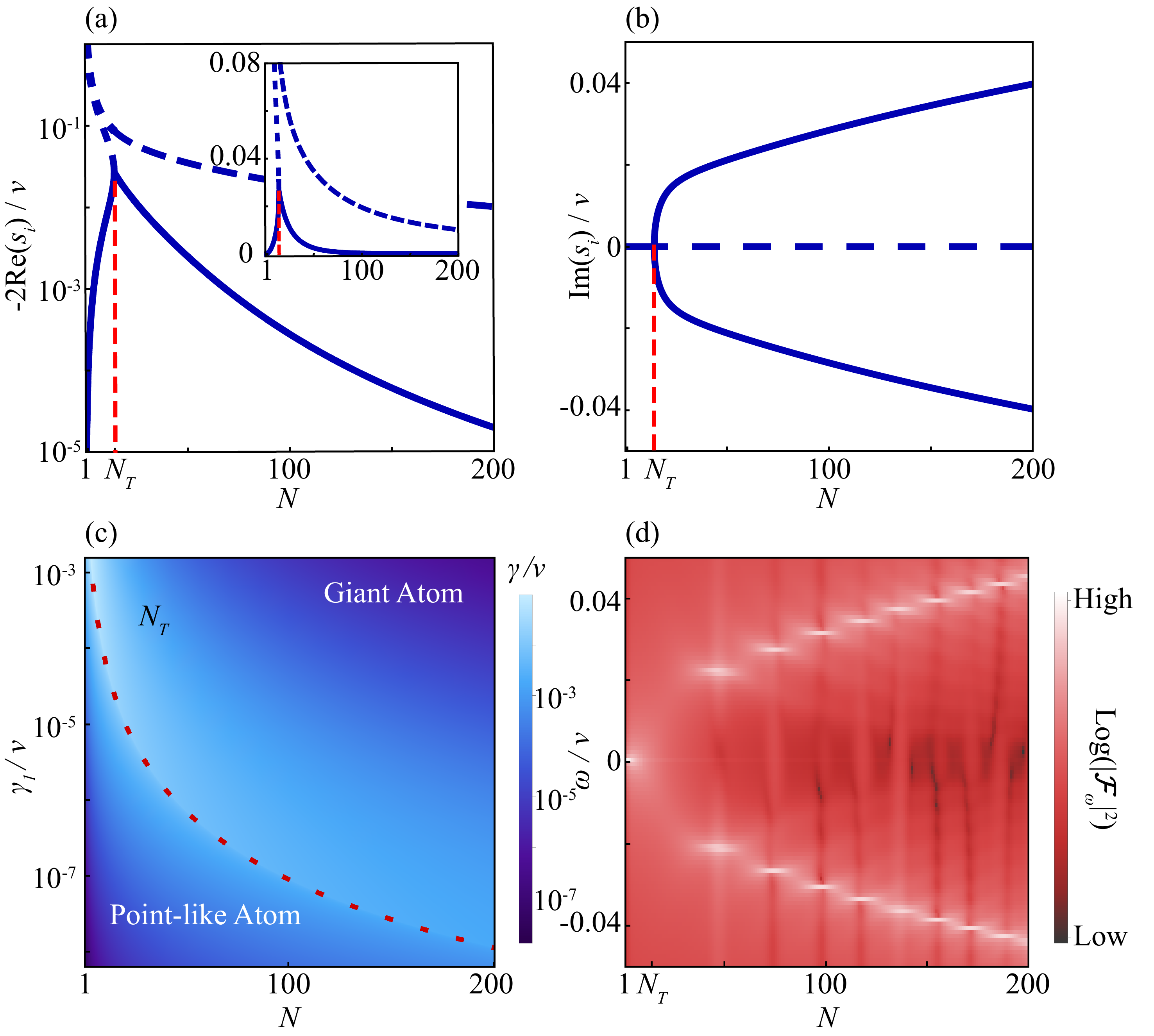}
        \caption{(a-c) Transition from the point-like atom to the giant atom, with the Lorentzian toy model: (a) The dark blue lines represent decay rates $-2\text{Re}(s_n)$ versus atom size $N$ in the semi-log scale,  where $s_n$ are roots for Eq.\ \eqref{eqn:main}. The blue solid one is the effective relaxation rate $\gamma \equiv  \text{Min}_n [-2\text{Re} (s_n)]$. The red dotted line shows  the transition point $N_T$. The inset is plotted in a linear scale. (b) The effective Rabi oscillation frequency $\text{Im}(s_n)$. (c) The effective relaxation rate $\gamma$ in the $N$-$\gamma_1$ parameter plane. The red dashed line shows  the transition point $N_T$, which separates two regimes for point-like atom and giant atom. (d) The power spectrum $|\mathcal{F}_\omega[\alpha(t;N)]|^2$ of the simulated time evolution with the circuit QAD model, in the log scale. We note that the discrete resonances observed arise from the phase matching condition in the circuit QAD model that is absent in the Lorentzian model.}
        \label{fig:3}
    \end{figure}    
    
    In Fig.\ \ref{fig:3}(a-b), we observe a dramatic change of dynamics at the \textit{transition point} $N_T$. When $N \ll N_T$, increasing the atom size only creates a larger coupling region and therefore accelerates the relaxation process. And at the transition point $N = N_T$, we find the imaginary parts of two poles merge, while their real parts split. And when $N \approx N_T$, the atom decays quickly into the 1-D waveguide, as all the modes have large damping rates. However, when $N > N_T$, the effective relaxation rate $\gamma$ drops almost exponentially with $N$, while the effective Rabi frequency becomes non-zero and increases. This result shows that a bounded giant atom regime exists at $N \gg N_T$, where some of the atomic excitation energy is localized and oscillates between atomic excitation and a stationary phonon wave packet. We also find that in the limit $N\to \infty$, Eq.\ \eqref{eqn:main} reduces to $s_n \to \pm (i/2)\sqrt{N\gamma_1\nu}$. As $\text{Re}(s_n) \to 0$, a part of the excitation lives in bound states in this limit. We can derive the transition point $N_T$ from the roots of Eq.\ \eqref{eqn:main}:
    \begin{equation}
        N_T = \sqrt[3]{\frac{(5\sqrt{5}-11)\nu}{2\gamma_1}} \approx 0.448 \times \sqrt[3]{\nu/\gamma_1}.
        \label{eqn:nt}
    \end{equation}
    For $\gamma_1 = \pi\times 10^{-5}\nu$, we have $N_T \approx 14.2$. In Fig.\ \ref{fig:3}(c), we show the effective relaxation rate $\gamma$ in the $N$-$\gamma_1$ parameter plane. We find two slow relaxation regions corresponding to the point-like atom case and the bounded giant atom case, which are on either side of $N_T$.
    
\subsection{Numerical Results from the Circuit QAD Model}
    
    \begin{figure}[t]
        \includegraphics[width=\linewidth]{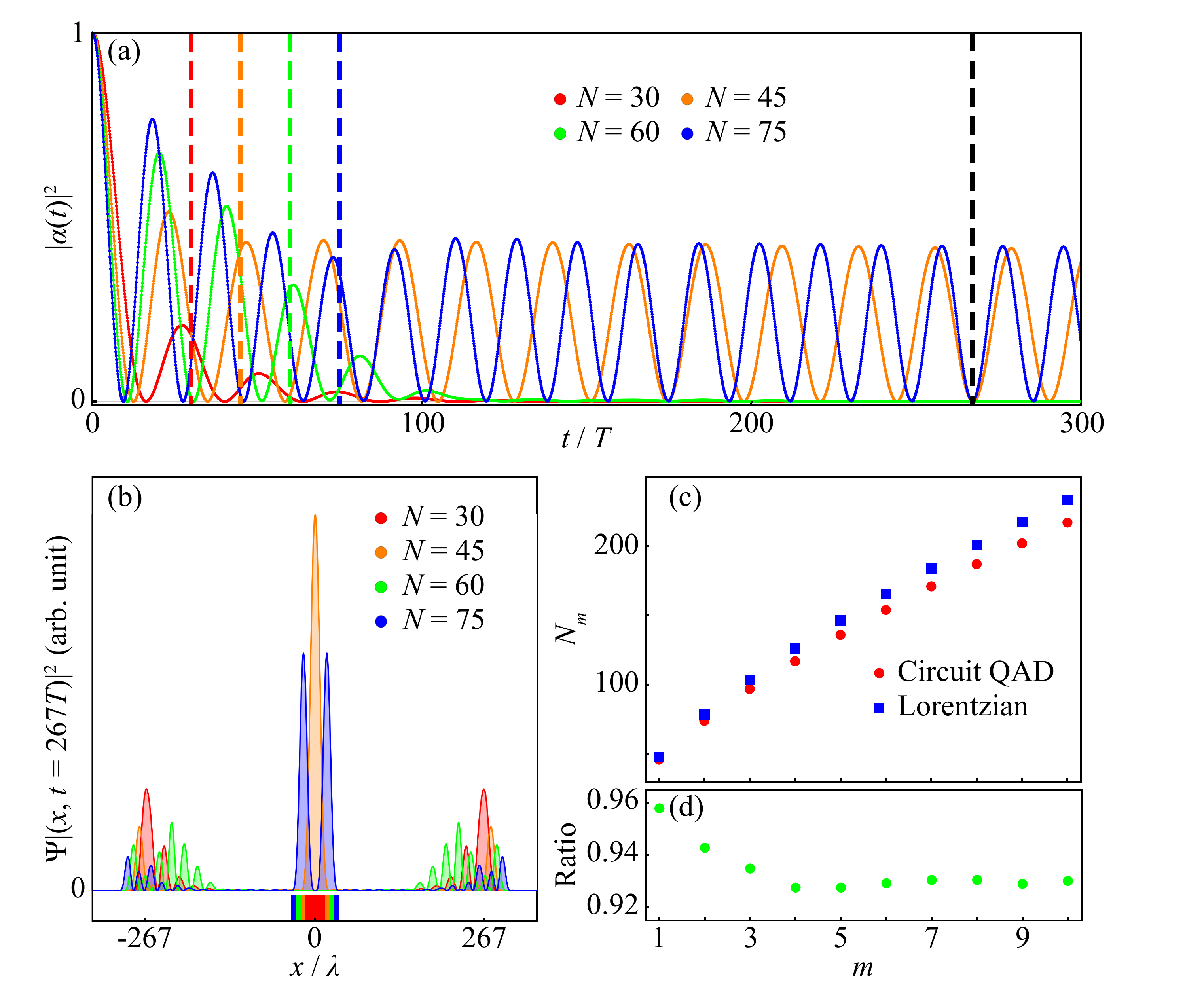}
        \caption{(a-b) Simulation for the circuit QAD model for different atom sizes $N$: (a) The time evolution of atom excitation amplitude $|\alpha(t)|^2$, for different $N$. The dashed lines show the time that the phonons travel through the atom length $t_b = NT$. For $N$ above the onset of normal mode splitting and phase matched, the system settles into a long-lived state after a short time. (b) The magnitude of phonon wave function $|\Psi(x,t_f)|^2$ frozen at $t_f = 267T$ (also indicated by the black dashed line on Fig.\ \ref{fig:4}(a)),  for different $N$. We chose $t_f$ such that $|\alpha(t_f)^2|\approx 0$ for all $N$ values shown. The colored box under $x$ axis represents the size of the atom. (c) The Lorentzian theory prediction and the circuit QAD simulation result of the bounded atom size $N_m$. (d) The ratio between $N_{m,\text{cQAD}} / N_{m,\text{Lor}}$.}
        \label{fig:4}
    \end{figure}
    
    Although it is hard to analytically evaluate the integral in Eq.\ \eqref{eqn:response} with the circuit QAD model, we can discretize the Hamiltonian and simulate the dynamics of the system via solution of the Schrodinger equation for the case of a single initial excitation, i.e., $\ket{\psi(t=0)} = \ket{e,0}$. We choose the cutoff momentum $k_c = \pm 0.1\pi/\lambda$ and the density of states $dk = 2\pi\times 10^{-4}/\lambda$, and time step $dt = 0.1 T$. We keep $\gamma_1 = \pi\times 10^{-5}\nu$ to compare with analytic results from the last subsection.
    
    In Fig.\ \ref{fig:4}(a), we show the time evolution of the atomic excitation, $|\alpha(t)|^2$. As expected, we find that for some $N\gg N_T$, such as $N=45$, and $75$, a fraction of the energy remains in the system after the phonons travel through the atom, i.e., $t_b = N T$, and this energy oscillates between mechanical and atomic excitation. Next, we choose a final time $t_f$, such that $|\alpha(t_f)|^2\approx 0$ for all the $N$ values we chose, and plot the magnitude of the phonon wavefunction $|\Psi(x,t_f)|^2$ in Fig.\ \ref{fig:4}(b). Again, we find that for $N = 45$, and $75$, a portion of energy remains confined within the range of the IDT after a long time.  
    
    We also show the logarithm of the power spectrum $|\mathcal{F}_\omega[\alpha(t;N)]|^2$ in Fig.\ \ref{fig:3}(d), where $\mathcal{F}_\omega[f(t)]$ represents the Fourier transform of $f(t)$. We observe qualitative agreement between Fig.\ \ref{fig:3}(b) and \ref{fig:3}(d) in terms of the locations of peaks when peaks are observed, but with discrete frequencies rather than continuous as a function of $N$. For example, from Fig.\ \ref{fig:4}(a-b), we also find that for some other $N\gg N_T$, such as $N = 60$, the atom still decays fast into the continuum and no peak is seen in the power spectrum. This behavior is caused by a mismatch between the atom length $N\lambda$ and the effective ``vacuum Rabi wavelength'' $\lambda_R(N) = 2\pi c_s/\text{Max}_n[\text{Im}(s_n(N))]$, as the circuit QAD model introduces a hard spatial boundary to the atom. Therefore, the circuit QAD model requires the atom size $N_m$ to satisfy an additional phase matching condition $N_m\lambda \approx m\lambda_R(N_m)$ for the bounded giant atom phenomenon, where $m\in \mathbb{N}$. We have discussed the first two cases, $N_1 = 45$ and $N_2 = 75$, and we further observe $1\sim 2$ peaks that correspond to $m$ in Fig.\ \ref{fig:4}(b). In Fig.\ \ref{fig:4}(c), we show a comparison between a  numerical simulation of the circuit QAD model (by finding largest resonances on the power spectrum, i.e. the brightest points on Fig.\ \ref{fig:3}(d)), and analytic calculations of $N_m$ using the Lorentzian model (by solving the equation $N_m\lambda = 2\pi c_s m/\text{Max}_n[\text{Im}(s_n(N_m))]$). Again, we find a qualitative agreement between two models. We also plot the ratio between $N_{m,\text{cQAD}}$ and $N_{m,\text{Lor}}$, which is stabilized around $0.93$ for $m\geq 3$. 
    
\section{Discussion and Conclusion}

    In this work, we have generalized the Weisskopf-Wigner theory from a point-like atom to a bounded giant atom that interacts with the medium instantaneously over a continuous spatial length $N \lambda$, with a simple Lorentzian toy model. When the coherence of the atom travels through the antenna much faster than the emission, we have observed that if its size $N$ satisfies both (1) the atom size $N$ is larger than the transition size $N_T$ and (2) the phase matching condition $N\lambda \approx m \lambda_R(N)$, a giant atom dynamic emerges, which is characterized by suppressed relaxation and effective vacuum Rabi oscillation with a phononic wave packet bound to the antenna, even in the absence of a cavity. To verify our results, we have compared it with the exact numerics of a realistic circuit QAD coupling model.   We have specifically studied the circuit QAD apparatus, but our analysis can be applied similarly to other quantum electro-mechanical systems with a large coupling spatial range \cite{Tom2019,Kim2017,coldatom}. For example, an optomechanical system where a membrane and a microwave waveguide coupled via radiation pressure could have similar effects.
    
    \textit{Note added:} During the preparation of this work, we learned of a similar result in Ref.~\cite{LZGuo2019}.
    

\section{Acknowledgement}

    We thank D. Carney, C. Flowers, J. Kunjummen, F. Liu., Y. Nakamura, A. Noguchi, K. Sinha,  E. Tiesinga, A. Gorshkov, and K. Srinivasan for insightful discussions. Y.W. acknowledges support by ARL CDQI, ARO MURI, NSF PFC at JQI, AFOSR, DoE BES QIS program (award No. DE-SC0019449), DoE ASCR Quantum Testbed Pathfinder program (award No. DE-SC0019040), DoE ASCR FAR-QC (award No. DE-SC0020312), and NSF PFCQC program. 


\begin{widetext}    
\section*{Appendix A: Derivation of the Circuit QAD Model}
    
    Consider the system described by Fig.\ \ref{fig:1}, where the IDT aligns to the [110] direction of a cubic crystal substrate. We assume the electrodes of the IDT do not change the mass density on the surface, and we model the Josephson junction as an LC circuit with inductance $L_J$ and capacitance $C_J$. The Lagrangian of the system is \cite{SAWbook}
    \begin{equation}
    \begin{split}
    \mathcal{L} &= \frac{L_J}{2}\dot{Q}^2 - \frac{1}{2C_\Sigma}Q^2 +  \frac{W}{2}\int_0^\infty dz  \int_{-\infty}^\infty dx \left[\rho (\dot{u_x}^2+\dot{u_z}^2)- c_{11}'(\frac{\partial u_x}{\partial x})^2 -  c_{11}(\frac{\partial u_z}{\partial z})^2-2c_{12}\frac{\partial u_x}{\partial x}\frac{\partial u_z}{\partial z} -  c_{44}(\frac{\partial u_x}{\partial z} + \frac{\partial u_z}{\partial x})^2\right]\\
    &- We_{14}\int_0^\infty dz\int_{-d/2}^{d/2} dx \left[\frac{\partial V}{\partial x}(\frac{\partial u_x}{\partial z} + \frac{\partial u_z}{\partial x})\right],
    \end{split}
    \label{eqn:Lag}
    \end{equation}
    where variables $Q(t)$ and $\vec{u}(x,z,t) = \{u_x,u_z\}(x,z,t)$ are the charge and strain degrees of freedom, respectively. The total capacitance $C_\Sigma = C_J+C_{\text{IDT}}$, where the capacitance of IDT $C_{\text{IDT}}$ can be calculated according to \cite{CIDT}. $W$ is the width of the IDT. The material parameters $\rho$, $c_{11}$, $c_{12}$, $c_{44}$,  $e_{14}$ are the density, elements of elastic tensor, and piezoelectric tensor of the substrate. For the cubic crystal, we have $c_{11}'=(c_{11}+c_{12}+2c_{44})/2$ \cite{stoneley1955}. To represent SAW modes, we take the ansatz \cite{stoneley1955}:
    \begin{align}
    u_x (x,z,t) = \sum_{j=-\infty}^{\infty}  C_j(t)\xi_j(z)\psi_j(x),
    \label{eqn:ux}\\
    u_z (x,z,t) = \sum_{j=-\infty}^{\infty}  C_j(t)\zeta_j(z)\psi_j(x),
    \label{eqn:uz}
    \end{align}
    where $\psi_j(x) = \sqrt{\frac{2}{L}}e^{-i K_j x}$, $\xi_j(z) = \sqrt{\frac{2}{L}}e^{-qK_j z-i\phi}$, and $\zeta_j(z) = \sqrt{\frac{2}{L}}r e^{-qK_jz-i\phi}$ with periodic boundary conditions in $x$, and $\vec{u}=0$ at $z\to \infty$. $L$, and $K_j = \frac{\pi j}{L}$ are the length of the system, and the momentum of modes, where $j\in \mathbb{Z}$. Th fitting parameters $q,r \in \mathbb{C}$, and $\phi\in\mathbb{R}$ can be derived from \cite{stoneley1955}. The electric field oscillates rapidly enough that the electric potential $V(x)$ is always quasi-static by the comparison of electron transmission. Therefore, we make the approximation:
    \begin{equation}
    \begin{split}
    \frac{\partial V}{\partial x} =
    \begin{cases}
    -\frac{2Q}{C_\Sigma\lambda}, &\text{for } \frac{2\eta-N}{2}\lambda\leq x< \frac{2\eta+1-N}{2}\lambda\\
    +\frac{2Q}{C_\Sigma\lambda}, &\text{for } \frac{2\eta+1-N}{2}\lambda\leq x< \frac{2\eta+2-N}{2}\lambda
    \end{cases},
    \label{eqn:v}
    \end{split}
    \end{equation}
    where $\eta=0,1,2, ...,N-1$. Substituting Eq.\ (\ref{eqn:ux}-\ref{eqn:v}) into Eq.\ \eqref{eqn:Lag}, we get
    \begin{equation}
    \mathcal{L} = \frac{L_J}{2}\dot{Q}^2 - \frac{1}{2C_\Sigma}Q^2 + \frac{W}{2L}  \sum_{j=-\infty}^{\infty} \left[  \frac{\rho'}{K_j}|\dot{C}_j|^2 -c' K_j |C_j|^2-\frac{e'}{C_\Sigma} \frac{  \sin
       \left(\frac{K_j \lambda  N}{2}\right)\tan \left(\frac{K_j \lambda }{4}\right)}{K_j\lambda}  Q C_j\right].
    \end{equation}
    
    The new parameters $\rho'=\rho(1+|r|^2)/\text{Re}[q],\ c' = \{c_{11}'+ c_{44}|r|^2 +(c_{44} +  c_{11}|r|^2)|q|^2 + i[c_{12}(r^* q^*-r q)+ c_{44} (r q^*-r^* q)]\}/\text{Re}[q]$, and $e' = 8e_{14} \text{Re}[(i-\frac{r}{q}) e^{- i\phi}]$ are effective density, elastic constant and piezoelectric constant, respectively. Then we define the momentum conjugates as: $V = L_J \dot{Q}$, $P_j= M_j \dot{C}_j$, where $M_j=\frac{W\rho'}{L K_j}$. And then we can calculate the quantized Hamiltonian by mapping $C_j \to \sqrt{\frac{\hbar}{2M_j\omega_j}}(\hat{a}_j+\hat{a}_j^\dagger),\ P_j \to -i\sqrt{\frac{\hbar M_j\omega_j}{2}}(\hat{a}_j-\hat{a}_j^\dagger),\ Q \to \sqrt{\frac{\hbar}{2L_J\nu}}(\hat{\sigma}_-+\hat{\sigma}_+),\ V \to -i\sqrt{\frac{\hbar L_J\nu}{2}}(\hat{\sigma}_--\hat{\sigma}_+)$. Then we have
    \begin{equation}
    \hat{H} = \hbar \nu \hat{\sigma}_+\hat{\sigma}_- + \sum_{j=-\infty}^{\infty} \hbar \omega_j\hat{a}_j^\dagger\hat{a}_j + \frac{\hbar g_0}{\sqrt{L/\pi}} \sum_{j=-\infty}^{\infty} \frac{  \sin
       \left(K_j \lambda  N/2\right)\tan \left(K_j \lambda/4\right)}{K_j \lambda/\pi}(\hat{\sigma}_-+\hat{\sigma}_+)(\hat{a}_j+\hat{a}_j^\dagger), 
    \end{equation}
    where $\nu \equiv 1/\sqrt{L_J C_\Sigma}$, $\omega_j \equiv c_s K_j$ (and $c_s=\sqrt{c'/\rho'}$), and $g_0 \equiv e' \sqrt{\frac{\pi W \nu}{ C_\Sigma\sqrt{c'\rho'} }}$. Taking the rotating wave approximation, the limit $L\to\infty$ then moving in to the rotating frame, we get the Hamiltonian Eq.\ \eqref{eqn:hamiltonian} with Eq.\ \eqref{eqn:cqad}.
    
    Then take the parameters of GaAs \cite{Schuetz2015} to estimate $\gamma_1$: $c_{11} = 12.26,\ c_{12}=5.71,\ c_{44}=6.00,\ c'_{11} = 14.99$ ($\times 10^{10}\text{N/m}^2$),$\ q=0.5+0.48 i,\ r = -0.68+1.16 i,\ \phi=1.05$, $\rho = 5307 \text{kg/m}^3$, $e_{14}= 0.157\text{C/m}^2$, and assume reasonable parameters as $\nu = 5\text{GHz}$, $C_\Sigma = 2.5\times10^{-11}\text{F}$, $W=50 \mu\text{m}$. Then our numerical estimations of parameters are: $\rho' = 14902\text{kg/m}^3,\ c'= 28.73 \times 10^{10}\text{N/m}^2,\ e' = -1.248 \text{C/m}^2,\ g_0 = -19.34 \sqrt{\mu\text{m}}\text{MHz},\ c_s = 4391 \text{m/s}$. $\gamma_1 =4\pi g_0^2/c_s\approx 0.34\pi \text{MHz} = 6.8\pi\times 10^{-5} \nu$. 
    
\section*{Appendix B: Explicit Solutions of the Lorentzian Model}
    
    Here we provide the explicit form for the roots of Eq.\ \eqref{eqn:main}:
    \begin{align}
        s_n = -\frac{\nu }{3 N}+ \frac{e^{-(2i\pi/3)n } \nu  \left(\nu -3 \gamma _1 N^3\right)}{6 A}+\frac{A e^{(2i\pi/3)n }}{6 N^2}
    \end{align}
    where $n = 1,2,3$, and $A = \sqrt[3]{-18 \gamma _1 \nu ^2 N^6+\nu ^3 N^3+3 \sqrt{3} \sqrt{\gamma _1 \nu ^3 N^9 \left(\gamma _1^2 N^6+11 \gamma _1 \nu  N^3-\nu ^2\right)}}$. We can find the transition point $N_T$ by take the square root part of $A$ equals zero, i.e. $\gamma _1^2 N_T^6+11 \gamma _1 \nu  N_T^3-\nu ^2 = 0$.
    
\section*{Appendix C: Top-hat Model and Bound States in Continuum}
    If $\gamma = 0$, then there exists at least one bound state in the 1-D continuum. Such a state is known as a bound state in continuum (BIC) \cite{BICrev,BICrev1975,nonM-BIC} or a decoherence-free state\cite{Anton2018,DFree1,DFree2,DFree3}. A BIC is an eigenstate of the Hamiltonian with eigenenergy within the continuum of the spectrum. Its existence usually requires symmetry protection or fine-tuning \cite{BICrev}. We illustrate the bound state in the continuum using the top-hat toy model
    \begin{equation}
        g_{\text{TH}}(k; N) \equiv \begin{cases} 
            \sqrt{\frac{\gamma_1 c_s}{2\pi}}\ N & |k|\leq \frac{2\pi}{N\lambda} \\
            0 & |k| > \frac{2\pi}{N\lambda} 
       \end{cases}.
        \label{eqn:TH}
    \end{equation}
    Note that though this toy model may seem simple, it is unphysical as it requires infinite spatial extent. Here, we report that a pair of purely imaginary solutions exist in our top-hat toy model. With Eq.\ \eqref{eqn:TH}, we can write the equation $[\chi(s_n)]^{-1}=0$ as
    \begin{equation}
        \pi s_n + i N^2\gamma_1 \log\left(\frac{N s_n - i \nu}{N s_n + i \nu}\right) = 0,
        \label{eqn:THeqn}
    \end{equation}
    where the complex function $\log(z)$ is the multiple-valued. Now we seek for purely imaginary solution $s_n = i\omega_n$, and we separate the real and imaginary part of Eq.\ \eqref{eqn:THeqn}, which results in
    \begin{align}
        2\pi \omega_n + N^2 \gamma_1 \log\left[\left(\frac{\nu-N \omega_n} {\nu+N\omega_n}\right)^2\right] = 0,\quad \text{with } |\omega_n| >\frac{\nu}{N}.
        \label{eqn:THBIC}
    \end{align}
    
    Although Eq.\ \eqref{eqn:THBIC} is transcendental, there always exists a pair of solutions for all $N$: We define the left-hand side of Eq.\ \eqref{eqn:THBIC} as $f(\omega_n)$, when $\omega_n\to \pm \nu/N$, $f(\omega_n)\to \mp\infty$; when $\omega_n\to \pm\infty$, $f(\omega_n)\to \pm\infty$. As $f$ is analytic, there exist a $\omega_1 < -\nu/N$ and a $\omega_2 > \nu/N$, such that $f(\omega_n)=0$.
\end{widetext}    
\bibliography{GiantAtom}
\end{document}